\documentclass[useAMS]{mn2e}
\usepackage{epsfig}
\title[Raman Scattered He~II~$\lambda$~4850 in V1016 Cyg]
{Centre Shift of the Raman Scattered He~II~$\lambda$~4850 in the
  Symbiotic Star V1016~Cygni}
\author[Jung \& Lee]{Yang-Chan Jung \thanks{E-mail:
ycjung@arcsec.sejong.ac.kr} and Hee-Won Lee 
\thanks{E-mail: hwlee@sejong.ac.kr} \\
Department of Astronomy and Space Science \\ 
Astrophysical Research Center for the Structure and Evolution
of the Cosmos \\ 
Sejong University, Seoul, 143-747, Korea }

\begin{document}

\date{Accepted 1988 December 15. Received 1988 December 14; in original 
form 1988 October 11}

\pagerange{\pageref{firstpage}--\pageref{lastpage}} \pubyear{2004}

\maketitle

\label{firstpage}

\begin{abstract}

We present our spectroscopic data around H$\beta$ of the symbiotic
star V1016~Cyg obtained with the Bohyunsan Echelle
Spectrograph, in order to secure the broad emission feature at around 4850
\AA,  which is formed through Raman scattering of He~II~$\lambda$~972.  
The total cross section around Ly$\gamma$ is approximately given by
$\sigma(\lambda)\simeq 1.7\times 10^{-28}[\lambda_{Ly\gamma}/(\lambda
-\lambda_{Ly\gamma})]^2{\rm\ cm^2}$, 
with $\lambda_{Ly\gamma}$ being the line centre wavelength of Ly$\gamma$.
We find a centre shift
redward by an amount $\Delta\lambda= +0.64{\rm\ \AA}$ in the Raman 
scattered He~II~$\lambda$~4850.  This redward centre shift is
exactly analogous to the effect for the Raman scattered
He~II~$\lambda$~6545 blueward of H$\alpha$ discussed in our previous
study.  We compute the branching ratios of Raman scattering into the
level $2s$ and levels $3s,$ and $3d$, which are subsequently incorporated in 
our Monte Carlo code.  Using this code, we present the centre shift of 
the 4850 feature as a function of the neutral hydrogen column density 
in the scattering region.  Assuming that He~II~$\lambda$~972 emission
is characterized by a single Gaussian profile, the redward peak shift 
observed in the Raman scattered He~II~4850 feature 
corresponds to the neutral column density $N_{HI}=1.2\times 
10^{21}{\rm\ cm^{-2}}$.
Assuming that the covering factor $\sim 0.1$ of the scattering region with
respect to the He~II emission region and adopting a simple spherical stellar
wind model, we may place an upper bound $\dot M \le 3.6\times 10^{-7}
{\rm\ M_\odot\ yr^{-1}}$ for the mass loss rate of the giant
component of V1016~Cyg. 
Our estimate can be severely affected by the kinematics of the scattering 
and He~II emission regions and the exact atomic physics, about which
brief discussions are presented.  
\end{abstract}


\section{Introduction}

Symbiotic stars, believed to be a binary system of a giant and a hot
white dwarf, are important objects for studying mass loss processes
occurring in the late stage of stellar evolution (e.g. Kenyon 1986). 
A thick neutral 
hydrogen region formed around the giant component as a result of the
mass loss is illuminated by strong UV radiation originating from the
hot white dwarf component. Schmid (1989) identified the broad emission 
features around 6825 \AA\ and 7088 \AA\ found in more than half of the
symbiotic stars by suggesting that they are formed via Raman scattering
of the resonance doublet O~VI~1032, 1038 by atomic hydrogen.
Simultaneous observations in the far UV regions and many
spectropolarimetric observations strongly support this identification
(e.g. Birriel, Espey \& Schulte-Ladbeck 1998, 2000, Schmid \& Schild 1994,
Harries \& Howarth 1996).

Raman scattering also appears to operate in the formation of 
H$\alpha$ wings that have been observed in most symbiotic stars 
and in a number of young planetary nebulae 
(Nussbaumer, Schmid \& Vogel 1989, Lee 2000, Lee \& Hyung 2000, Arrieta \&
Torres-Peimbert 2003). The H$\alpha$ profiles in the far wing regions 
are well fitted by
$f_\lambda\propto \Delta\lambda^{-2} =(\lambda-\lambda_{H\alpha})^{-2}$, 
which delineates approximately the Raman scattering cross section,
where $\lambda_{H\alpha}$ is the line centre wavelength of H$\alpha$.
He~II $n\rightarrow 2$ emission lines arising from states 
with even principal quantum numbers have wavelengths only a
little bit shorter than those of the H~I Lyman series lines. 
These He~II emission lines can be Raman scattered to
form broad features blueward of the Balmer series lines. 

We expect that the strongest Raman scattered He~II feature is formed
at around 6545 \AA\ blueward of H$\alpha$, which is formed through 
Raman scattering of
He~II~$\lambda$~1025. This feature was
found in the spectra of the symbiotic stars RR~Tel, He 2-106 and 
V1016~Cyg (Lee, Kang \& Byun 2001, Lee et al. 2003). Another broad feature
is formed at around 4850 \AA, which is the Raman scattered feature of 
He~II~$\lambda$~972. This feature was reported in  spectra of the
symbiotic stars RR Tel (Van Groningen 1993) and V1016~Cyg (Birriel
2003). It is quite notable that the young
planetary nebula NGC~7027 also exhibits this 4850 broad feature  
(P\'equignot et al. 1997). 

Raman scattered He~II lines can be used as a powerful diagnostic of the mass
loss processes in symbiotic stars, because
the mass loss rate can be calculated by inferring the optical depth
structures and the covering factor of the scattering region with
respect to the UV emission region around the white dwarf. However,
the flux of a Raman scattered He~II feature is mainly determined by the
product of the optical depth and the covering factor, which is a big
barrier for the exact estimate of the mass loss rate.

In the previous theoretical study by Jung \& Lee (2004, hereafter JL04), 
they showed that the Raman scattered He~II~$\lambda$~6545  should exhibit
a redward centre shift by a significant amount depending on the H~I column
density of the scattering region around the giant in the absence of
the relative motion between the scattering region and the emission
region, which will lift
the degeneracy of the optical depth and the covering factor. However, Raman
scattered He~II~6545 is usually blended with the forbidden line
[N~II]~$\lambda$~6548, making it very hard to determine the exact peak
location. This difficulty may be avoided when we use the Raman
scattered He~II~4850, which should be much weaker than the 6545
feature but be much easier to locate the peak wavelength. 

UV radiation around Ly$\gamma$ may have four scattering
channels depending on the final state of the scattering atom, which
can be one of $1s, 2s, 3s$ and $3d$. When scattering occurs into
either $3s$ or $3d$ level, energy conservation leads to the wavelength 
of the scattered radiation 
$\lambda_{3s, 3d}=(\lambda_{He~II~972}^{-1}-\lambda_{Ly\beta}^{-1})^{-1}
=1.85\ \mu$, where $\lambda_{He~II~972}$ and $\lambda_{Ly\beta}$
are line centre wavelengths for He~II~$\lambda$~972 and Ly$\beta$.
Therefore, in order to simulate
the Raman scattered He~II lines accurately, it is necessary to compute the branching
ratios into $3s$ and $3d$ states. This calculation will also be useful for
the future study of Raman scattered lines formed blueward of Paschen
series lines. 

In this paper, we present an optical spectrum around H$\beta$ of 
the symbiotic star
V1016~Cygni obtained with the 1.8~m Bohyunsan telescope and our Monte
Carlo calculations of Raman scattering around Ly$\gamma$, 
in order to determine the neutral hydrogen column density of the
scattering region around the giant component. 

\section{Observational Data}

\begin{figure}
 \vspace{12pt}
 \hspace{-20pt}
 \epsfig{file=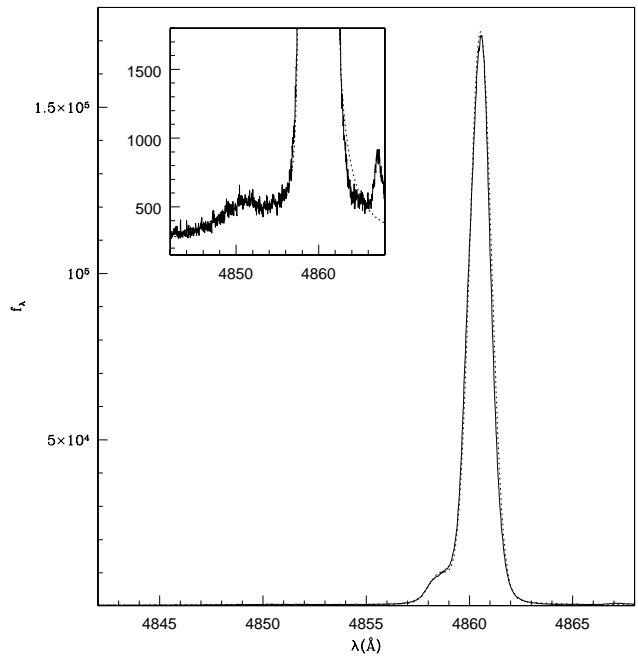, height=9cm, width=9cm}
 \caption{
The optical spectrum around H$\beta$ of V1016~Cyg obtained 
with the Bohyunsan Optical Echelle Spectrograph (BOES) installed on the
1.8~m telescope at Mt. Bohyun. The spectroscopic resolution is 44,000
and the exposure time was 3000~s. The thin solid lines represent
our single Gaussian fits to the emission lines H$\beta$, He~II~$\lambda$~4861
and the Raman scattered He~II~$\lambda$~4850. From our Gaussian fitting,
the Raman scattered He~II~$\lambda$~4850 is shifted redward by an amount
$\Delta\lambda = +0.64{\rm\ \AA}$ with respect to the emission line center
He~II~$\lambda$~4861.}
\end{figure}

We observed V1016~Cyg on the night of 2003 December 16 with the
Bohyunsan Optical Echelle Spectrograph equipped with a 2k$\times$4k
E2V CCD installed on the 1.8~m telescope at Mt. Bohyun. 
We used the 200~$\mu$ optical fiber that yields the spectroscopic resolution 
$R=44,000$ and the exposure time was 3000~s. The data have been
reduced following standard procedures using the IRAF packages.
In Fig.~1, we show the part of our spectrum that contains H$\beta$.
The solid line shows the observed spectrum, and in the inset
we enhanced the part exhibiting the Raman scattered He~II~$\lambda$~4850
feature.  In this figure, the vertical axis is shown in an arbitrary unit, 
which is sufficient for the purpose of this work.

We measure the relative shift from the emission lines
He~II~$\lambda$~4861 and H$\beta$, where these emission lines are
assumed to be isotropic in their emission region. This assumption may
not be true in the accuracy required in this work. In the case of
O~VI~$\lambda$~1032, 1038 doublet, it is known that their profiles and
those of their Raman scattered 6825, 7088 features differ quite
significantly (Schmid et al. 1999). The O~VI  doublet lines are strong
resonance lines that are subject to various radiative transfer
effects, whereas optical He~II recombination lines are expected to be
relatively free from such effects. This may be checked by monitoring
the variations as a function of the orbital phases. In this work, we
take the He~II~$\lambda$~4861 and H$\beta$ as the local reference of
velocity and measure the line centre shift of the 
Raman scattered He~II~$\lambda$~4850. 

In Fig.~1, we applied single Gaussian fits to both
He~II~$\lambda$~4861 and H$\beta$ to locate their peak positions. The dotted
lines show our fit, which we can barely notice in the figure. This implies
that satisfactory fits are obtained for both emission lines.
For He~II~$\lambda$~4861, the fitting function is
$f_{He~II}= f_1e^{-[(\lambda-\lambda_1)/\Delta\lambda_1]^2}$
and for H$\beta$ it is
$f_{H\beta}= f_2e^{-[(\lambda-\lambda_2)/\Delta\lambda_2]^2}$,
where $f_1=165000, f_2=8200, \lambda_1=4860.54{\rm\ \AA}, 
\lambda_2=4858.57{\rm\ \AA}$ and 
$\Delta\lambda_1=\Delta\lambda_2= 0.75{\rm\ \AA}$.

The 4850 feature is also fitted with a single Gaussian 
\begin{equation}
f_{4850} = f_3e^{-[(\lambda-\lambda_3)/\Delta\lambda_3]^2},
\end{equation}
where the least square method yields $\lambda_3 = 4850.63{\rm\ \AA}$
and $\Delta\lambda_3=3.76{\rm\ \AA}$. 
We note that the width broadening is in excellent agreement 
with the theoretical expectation 
\begin{equation}
\Delta\lambda_3/\Delta\lambda_1 \simeq {4861/ 972} = 5.0.
\end{equation}

Using the atomic data 
compiled by van Hoof, the centre wavelength $\lambda_0$ in vacuum of the Raman scattered 
He~II~$\lambda$~972 is
\begin{equation}
\lambda_0 = (\lambda_{He~II~972}^{-1}-\lambda_{\alpha}^{-1})^{-1}
=4852.098{\rm\ \AA},
\end{equation}
where the vacuum wavelength of Ly$\alpha$ is $\lambda_{\alpha}=1215.67{\rm\
\AA}$ and that for He~II~$\lambda$~972 is $\lambda_{He~II~972}=972.112
{\rm\ \AA}$.  Considering the refractive index of air 
$n_{air}=1.000279348$,
the line centre of the 4850 feature appears 
at $\lambda_3^0 = 4850.743{\rm\ \AA}$ in air. Noting that the line
center of He~II~$\lambda$~4861 in our spectrum appears at 
$\lambda_{HeII}=4859.320{\rm\ \AA}$,
our spectrum of V1016~Cyg shows the redward centre shift $\Delta\lambda
=+0.64{\rm\ \AA}$ in the 4850 feature with respect to their emission region.

\section{Atomic Physics}

We compute the matrix elements necessary for computation of the
contribution to the scattering cross section of the levels $3s$ and
$3d$. The contributions of the levels $1s$ and $2s$ have been
presented in the literature (e.g. Saslow \& Mills 1967, 
Sadegpour \& Dalgarno 1992, Lee \& Lee 1997).

\begin{figure}
 \vspace{12pt}
 \hspace{-10pt}
 \epsfig{file=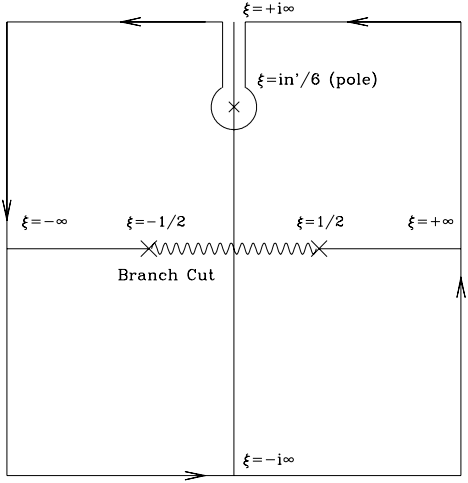, height=9cm, width=9cm}
 \caption{
The contour $C$ for the integral in Eq.~9. There is a branching cut
on the real axis between $\xi=1/2$ and $\xi=-1/2$. There is a pole
at $\xi = in'/6$, at which residues are computed.}
\end{figure}

According to Karzas \& Latter (1961), the matrix element for
bound-bound state transition is given by
\begin{eqnarray}
\tau_{nl}^{n' l-1} &=& \int_0^\infty r R_{n' l-1}(r) R_{nl}(r)  r^2 dr  
\nonumber \\
&=& {2^{2l}\over (2l-1)!} \left[ {(n+l)! (n'+l-1)! \over (n-l-1)!
    (n'-l)!} \right]^{1/2} (nn')^{l+1} \nonumber \\
&\times & (n+n')^{-n-n'} (n-n')^{n-2-l} (n'-n)^{n'-l}
\nonumber \\
&\times & \left\{ F(l-n+1,l-n',2l, x_{n,n'}) \right. \nonumber \\
&-&\left. \left({n-n' \over n+n'} \right)^2
F(l-n-1,l-n',2l, x_{n,n'}) \right\},
\end{eqnarray}
where $x_{n,n'}\equiv -4nn'/(n-n')^2$ and
$F(\alpha,\beta,\gamma,x)$ is the hypergeometric function 
\begin{equation}
F(\alpha,\beta,\gamma,x)\equiv  1+{\alpha\beta\over\gamma}{x\over 1!} 
+{\alpha(\alpha+1)\beta(\beta+1)\over \gamma( \gamma+1)}{x^2\over 2!} + \cdots.
\end{equation} 
Here, $R_{nl}(r)$ is the radial wavefunction of a hydrogen atom, which
can be expressed using a confluent geometric function $F_1$ by
\begin{eqnarray}
R_{nl}(r)&=&{1\over(2l+1)!}\sqrt{{(n+l)!\over(n-l-1)! 2n}}
\left({2Z\over n}\right)^{3/2} e^{-Zr/n} \nonumber \\
&&\left( {2Zr\over n}\right) F_1(-(n-l-1), 2l+2; 2Zr/n),
\end{eqnarray}
where $F_1(a,b;x)\equiv 1+abx+[a(a+1)b(b+1)/2!]x^2+\cdots$.

By setting $n'=3, l=1$, a straightforward calculation leads to
\begin{eqnarray}
\tau_{3s}^{np}&\equiv &<3s \parallel r \parallel np> 
={1\over 4}\left[{(n+1)! 3! \over (n-2)! (3-1)!}\right]^{1/2} \nonumber \\
&\times& {(12\ n)^2 (n-3)^{n+3-4}\over (n+3)^{n+3}}  \nonumber \\
&\times& \left\{ F(-n+2,-2,2,{-12n\over (n-3)^2}) \right. \nonumber \\
&-&\left.\left({n-3\over n+3}\right)^2 F(-n, -2, 2,{-12n\over (n-3)^2})
\right\} \nonumber \\
&=& 3^{3/2} 12^2 n^{7/2} \sqrt{n^2-1} (17n^2 - 27) { (n-3)^{n-4}\over
  (n+3)^{n+4} } . 
\end{eqnarray}

By interchanging $n$ and $n'$ in Eq.~(1) and letting $n'=3, l=2$, we
obtain
\begin{eqnarray}
\tau_{3d}^{np} &\equiv& <3d \parallel r \parallel np> 
{1\over 24}\left[{(n+1)! 5! \over (n-2)! 0!}\right]^{1/2} \nonumber \\
&\times& {(12n)^3 (n-3)^{n-3}\over (n+3)^{n+3}}  \nonumber \\
&\times& \left\{ F(0,-n+2,4,{-12n\over (n-3)^2})\right. \nonumber\\
&-&\left.\left({n-3\over n+3}\right)^2 F(-2, -n+2, 4,{-12n\over (n-3)^2})
\right\} \nonumber \\
&=& {12^3\sqrt{3}\over\sqrt{10}} n^{11/2} \sqrt{n^2-1} { (n-3)^{n-4}\over
  (n+3)^{n+4} } .
\end{eqnarray}

According to Bethe \& Salpeter (1957) the wavefunction for the
continuum states $n'p$ is given by the contour integral
\begin{eqnarray}
R_{n' l=1} &=& {[1+n'^2]^{1/2}\over \pi [1-e^{-2\pi n'}]^{1/2}}
{n'^2\over 4r^2} \nonumber \\
&\times &
\int_C e^{-2ir\xi/n'}(\xi+{1\over2})^{-in'-2}(\xi-{1\over2})^{in'-2} d\xi,
\end{eqnarray}
where the contour $C$ is depicted in Fig.~2.

The matrix element for a dipole operator between $3s$ and $n'p$ states is
\begin{eqnarray}
\tau_{n'p}^{3s} &=&<3s|r|n'p> \nonumber \\
&=&{n'^{1/2}[1+n'^2]^{1/2} \over 4\pi[1-e^{-2\pi n'}]}
\int_C (\xi+1/2)^{-in'-2}(\xi-1/2)^{in'-2} d\xi \nonumber \\
&\times& \int_0^\infty dr r e^{-{2i\xi r\over n'}}
{2\over 3\sqrt{3}}e^{-r/3}\left( 1-{2r\over3}+{2r^2\over27}\right).
\end{eqnarray}
Letting $\lambda ={1\over3}+{2i\xi\over n'}$ the radial integral becomes
\begin{eqnarray}
I_1 &=&{2\over 3\sqrt{3}}
\int_0^{\infty}e^{-r(1/3+2i\xi/n')}\left(
  1-{2r\over3}+{2r^2\over27}\right) dr \nonumber \\
&=&{2\over 3\sqrt{3}} \left[ {1\over\lambda^2}
  -{4\over3\lambda^3}+{4\over 9\lambda^4}
\right].
\end{eqnarray}
Using the residue theorem for the pole at $\xi=in'/6$, 
we compute the contour integral
\begin{eqnarray}
I_2 &=& \int_C d\xi (\xi+1/2)^{-in'-2}(\xi-1/2)^{in'-2}
\nonumber \\
&\times&\left[ {1\over\lambda^2}
  -{4\over3\lambda^3}+{4\over9\lambda^4}\right] \nonumber \\
&=&{6^5\pi n'^3 \over (n'^2+9)^4}
(7n'^2+27) e^{-2n'\tan^{-1}(3/n')}.
\end{eqnarray}
Substituting this result we have
\begin{equation}
\tau_{3s}^{n'p} = {3^{3/2}12^2 n'^{7/2} (1+n'^2)^{1/2}(7n'^2+27) \over
(1-e^{-2\pi n'})(n'^2+9)^4} e^{-2n'\tan^{-1}(3/n')}.
\end{equation}

A similar procedure gives
\begin{equation}
\tau_{3d}^{n'p} = {12^3\over \sqrt{10}} {n'^{11/2} (1+n'^2)^{1/2} \over
(1-e^{-2\pi n'})(n'^2+9)^4} e^{-2n'\tan^{-1}(3/n')}.
\end{equation}

\begin{figure}
 \vspace{12pt}
 \hspace{-10pt}
 \epsfig{file=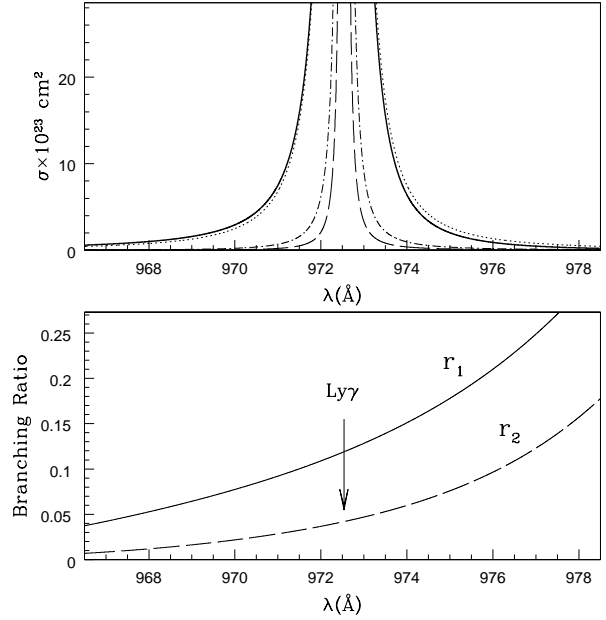, height=9cm, width=9cm}
 \caption{
 The scattering cross sections and branching ratios of UV radiation 
 around Ly$\gamma$ by atomic hydrogen. In the upper panel, the 
 solid line shows the total scattering cross section, the dot-dashed 
 line represents
 the scattering cross section into the level $2s$ (for He~II~$\lambda$~4850)
 and the dashed line represents the sum of the scattering
 cross sections into the levels $3s$ and $3d$. Also in the upper panel, 
 the dotted line shows
 a Lorentzian fit to the total scattering cross section by the formula
$\sigma_0[\lambda_{Ly\gamma}/(\lambda-\lambda_{Ly\gamma}]^{-2}$, where
$\lambda_{Ly\gamma}$ is the line centre wavelength of Ly$\gamma$ and
$\sigma_0=1.7\times10^{-28}{\rm\ cm}^2$. In the lower panel, we show the 
branching ratio $r_1$ of Raman scattering into the level $2s$ by a solid 
line, and the branching ratio $r_2$ into the levels $3s$ 
and $3d$ by a dashed line.
}
\end{figure}

In Fig.~3 we show the scattering cross sections and branching ratios 
as a function of the wavelength of UV radiation
around Ly$\gamma$. In the upper panel we show the total scattering
cross section $\sigma_T$ by a thick solid line, the cross section
$\sigma_{2s}$ for Raman scattering
into the $2s$ level by a dot-dashed line, and the cross section
$\sigma_{3s3d}$  for Raman
scattering into the $3s-3d$ levels by a dashed line.  

Also in the upper panel, the dotted line shows
a Lorentzian fit to the total scattering cross section,
which is given by
\begin{equation}
\sigma(\lambda)=\sigma_0
\left({\lambda_{Ly\gamma}\over \lambda-\lambda_{Ly\gamma}} \right)^2.
\end{equation}
Here, $\lambda_{Ly\gamma}$  is the line centre wavelength for the Ly$\gamma$
transition. With this fit, we set $\sigma_0 = 0.17\times 10^{-27}{\rm
cm^2}$. As is for the case of Ly$\alpha$ shown  by Lee (2003), the total
cross section is also asymmetrical with respect to the Ly$\gamma$ line 
center, which is noticeable in this fit.
At the line centre $\lambda = 972.112{\rm\ \AA}$ of He~II~$\lambda$~972, 
the total scattering cross section is
\begin{equation}
\sigma_{tot, He~II~972}=0.91\times 10^{-21}{\rm\ cm^2}.
\end{equation}
This may be compared with the total scattering cross section for
He~II~$\lambda$~1025, which is $\sigma_{tot, He~II~1025}=6.2\times 10^{-21}
{\rm\ cm^{-2}}$.
From this computation, we may regard the Raman scattered He~II~$\lambda$~6545
as a probe for neutral regions with $N_{HI}\sim 10^{20}{\rm\ cm^{-2}}$
and the Raman scattered He~II~4850 is formed mainly in regions with $N_{HI}
\ge 10^{21}{\rm\ cm^{-2}}$. 

In the lower panel of Fig.~3, we show the
branching ratios $r_1=\sigma_{2s}/\sigma_T,
r_2 = \sigma_{3s3d}/\sigma_T$ for Raman scattering into the $2s$ level
and the $3s-3d$ levels by a solid line and a dashed line, 
respectively.
It is quite notable that the branching ratio into the level $2s$ is
about 3 times larger than the sum of the branching ratios 
into $3s$ and $3d$ near He~II~$\lambda$~972. 
Raman scattering into levels $3s$ and $3d$ will leave
a broad feature at $\lambda=1.85 \mu$ blueward of Pa$\alpha$. According
to this calculation, it is expected that the $\lambda=1.85 \mu$ feature
will be constituted by a smaller number of photons by a factor of 1/3 
than those in the Raman scattered He~II~$\lambda$~4850.  
At around $\lambda = 985{\rm\ \AA}$ the total cross
section is dominated by the scattering channel into the levels $n=3$,
for which, however, the scattering cross section is very small. Near the
line center of Ly$\beta$ the branching ratio for the levels $n=3$
becomes negligibly small due to the decrease of the phase space volume
and becomes zero redward of Ly$\beta$, which is forbidden 
by energy conservation.

\section{Calculations}

\subsection{Monte Carlo Procedures}

The same Monte Carlo code used in the previous study by JL04
is slightly modified with an addition of subroutines for 
those scattering channels into $3s$ and $3d$ levels. 
The code incorporates the channel for Rayleigh scattering
as well as the two channels for Raman scattering. Therefore, we followed
faithfully paths of He~II photons suffering several Rayleigh scatterings
followed by a Raman scattering in the H~I region.  
However, no dust effect is considered in the current work.

As in JL04,
considering the fact that He~II~6560, He~II~4686 emission lines 
are well fitted  by a single Gaussian 
\begin{equation}
f_\lambda = f_0 \exp[-(\lambda-\lambda_0)^2/\Delta\lambda^2],
\end{equation}
with the velocity width $\Delta\lambda/\lambda_0 = (23{\rm\ km \
  s^{-1}}/c)$. We use the same single Gaussian for the incident
  radiation.  Since we are only concerned with the exact location 
  of the central
  peak, we do not require the exact equivalent width of the
  He~II~$\lambda$~972 line and set it to be $0.1{\rm \AA}$.

When the scattering region is optically thin, the strength of 
the Raman scattered feature
is mainly determined by the product of the incident radiation flux and
the scattering cross section. The He~II~$\lambda$~972
incident line radiation has a profile that is symmetric with respect
to its line center, whereas the scattering cross section is highly
inclined toward the line center of Ly$\gamma$, which is located
redward of He~II~972. This induces a redward shift of the peak in the
Raman scattered He~II~$\lambda$~4850, as is explained for the Raman
scattered He~II~$\lambda$~6545 in JL04. We illustrate this situation 
in Fig.~4, where
the dotted line shows the Raman scattering cross section 
into the level $2s$ relevant
to the formation of the 4850 feature. In this figure, the dashed
line shows the sum of the Raman scattering cross sections 
into the levels $3s$ and
$3d$, which is slightly larger than the quantity represented by the
dotted line in this region.

Due to the multiple scattering effect that is important for
a very optically thick scattering region, the peak shift will decrease
as $N_{HI}$ increases, and therefore
a Monte Carlo technique is useful to obtain a reliable quantitative relation.

\begin{figure}
 \vspace{12pt}
 \epsfig{file=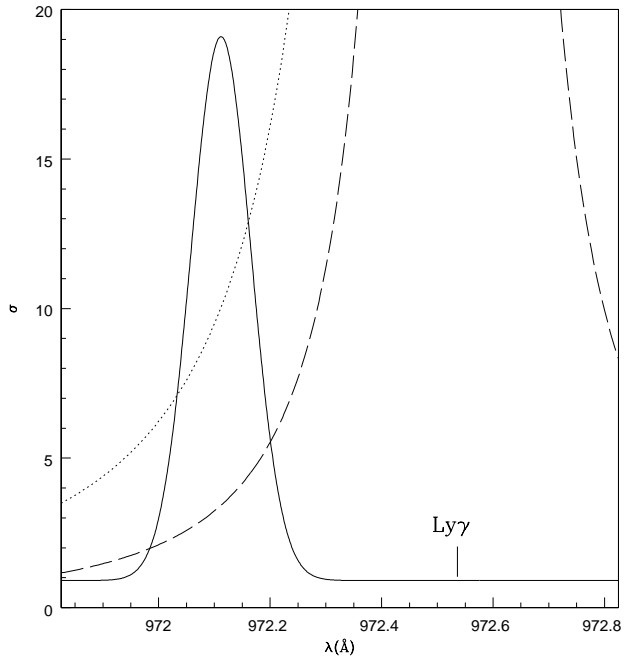, height=9cm, width=9cm}
 \caption{
The Raman scattering cross sections overplotted to the He~II~972
emission line given by a single Gaussian with $\Delta\lambda
=0.076{\rm\ \AA}$. The dotted line shows the cross
section of the Raman scattering into the $2s$ level, and the 
dashed line shows the sum of the cross sections for the Raman
scattering into the $3s$ and $3d$ levels.  
Note that the cross section increases steeply toward
the Ly$\gamma$ line center, which will affect the center location
of the Raman scattered He~II~4850 feature. }
\end{figure}

\subsection{Results}

\subsubsection{Balmer and Paschen Wings}

\begin{figure}
 \vspace{12pt}
 \hspace{-10pt}
 \epsfig{file=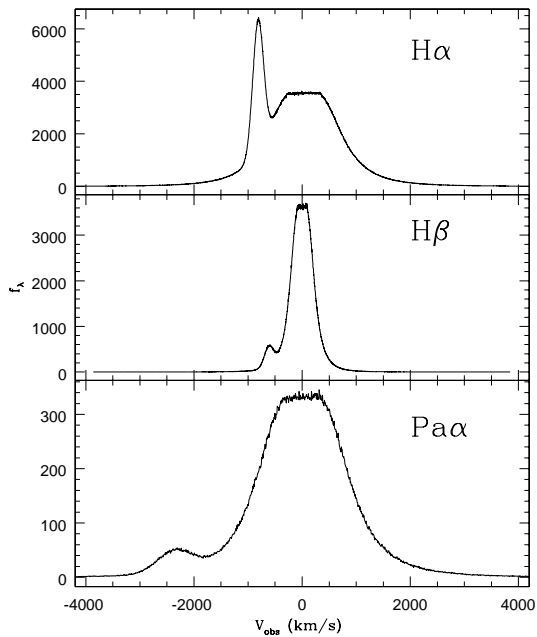, height=9cm, width=9cm}
 \caption{
Typical Raman scattered He~II features formed
blueward of H$\alpha$, H$\beta$ and Pa$\alpha$. The H~I column
density of the scattering region is assumed to be
$N_{HI}=3\times 10^{20}{\rm\ cm^{-2}}$.
We normalize the horizontal scale by the velocity
$c(\Delta\lambda_R/\lambda_R)$ from the line
centres of Raman scattering, where $\lambda_R=6563, 4861,$ and
$18751{\rm\ \AA}$ respectively.
Note that with this normalization the flux density is diluted by
the factor $\lambda_R/\lambda_{Ly\gamma}$.
The vertical axis shows the number of photons
obtained from the Monte Carlo simulations for $5\times 10^4$ 
continuum photons per
wavelength interval of $\Delta\lambda=0.00125{\rm\ \AA}$ in the UV rest frame
around Ly$\gamma$. 
}
\end{figure}

In Fig.~5, we show typical Raman scattered He~II features formed
in the blue region of H$\alpha$, H$\beta$ and Pa$\alpha$, where we set 
$N_{HI}=3\times 10^{20}{\rm\ cm^{-2}}$. The
formation of the 6545 feature was studied intensively by JL04, and
here it was reproduced for comparisons with those obtained from Raman
scattering of UV radiation around Ly$\gamma$. 
The vertical axis shows the number of photons
obtained from the Monte Carlo simulations for $5\times 10^4$ continuum
photons per wavelength interval of $\Delta\lambda=0.00125{\rm\ \AA}$ in the
UV rest frame around Ly$\gamma$. 

We normalize the horizontal scale by the velocity
$c(\Delta\lambda_R/\lambda_R)$ from the line
centres of Raman scattering, where $\lambda_R=6563, 4861,$ and
$18751{\rm\ \AA}$ respectively. With this normalization, the flux
density at $\lambda_R$ in the Raman scattered features is diluted by the factor
$\lambda_R/\lambda_{Ly\gamma}$. 

Far from the line centres of Balmer lines
and Pa$\alpha$, the Rayleigh and Raman scattering optical depth is
quite small, and the Raman wings exhibit profiles that are proportional
to the total scattering cross section multiplied by the corresponding
branching ratio. 
It is notable that the sum of the branching ratios for levels $3s$ and $3d$
is smaller by a factor 1/3 than the branching ratio for level $2s$, and 
therefore our simulations show that the Raman scattered feature blueward of 
Pa$\alpha$ is weaker by the similar factor than the feature blueward 
of H$\beta$.

Near the line centres of Balmer lines and Pa$\alpha$, the total
scattering optical depth is quite large and the Raman conversion
efficiency no longer increases toward the line center. Instead, almost
all incident Ly$\gamma$ photons may be singly scattered at the surface of the scattering
region by a Rayleigh or Raman process, 
or may suffer a few Rayleigh scatterings followed
by Raman scattering or final Rayleigh scattering before escape, 
which leads to nearly constant Raman conversion efficiencies 
into around H$\beta$ and Pa$\alpha$. 
More quantitatively, according to our Monte Carlo results, 
about 36 percent of incident photons very near Ly$\gamma$ line
center are converted and redistributed around H$\beta$ and 13 percent of them
are redistributed around Pa$\alpha$. 

It is also quite notable that the Raman feature blueward of Pa$\alpha$
appears 4 times further away from the centre of Pa$\alpha$ than the
4850 feature is away from the H$\beta$ centre. This
is attributed to the difference in the ratio $\lambda_o/\lambda_i$ 
of the wavelengths
of the Raman scattered radiation and incident UV radiation.

From this figure, we infer that the
wings around H$\beta$ and Pa$\alpha$ are very weak compared with those
around H$\alpha$, unless the continuum around Ly$\gamma$ is
unrealistically stronger than that around Ly$\beta$. 
The relative strengths of H$\alpha$ and H$\beta$ wings may be useful
in distinguishing those wings that are formed from other mechanisms
including Thomson scattering.

\subsubsection{H$\beta$ wings and the Raman Scattered He~II~4850}

\begin{figure}
 \vspace{12pt}
 \hspace{-10pt}
 \epsfig{file=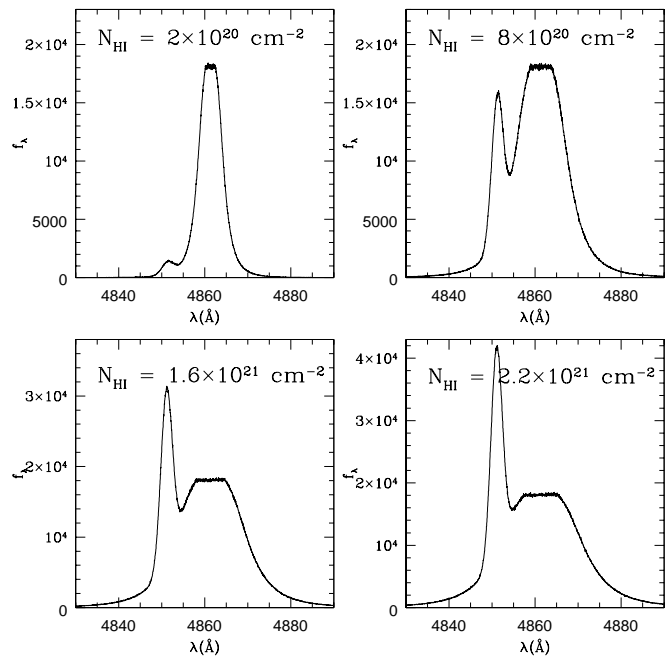, height=9cm, width=9cm}
 \caption{
The Raman scattered He~II~$\lambda$~4850 feature formed in neutral
scattering regions with various H~I column densities. The strength
and the centre shift of the He~II~$\lambda$~4850 feature exhibit
similar behaviours to those of the He~II~$\lambda$~6545 feature
discussed by JL04. The incident radiation is prepared in the same
way as in Fig.~5 and the vertical axis has the same meaning as in Fig.~5.
}
\end{figure}

In Fig.~6 we show H$\beta$ wings and the Raman scattered He~II~$\lambda$~4850
feature for 4 values of $N_{HI}$. The extent of H$\beta$ wings increases
as $N_{HI}$ increases and the wing profile is approximately by
$(\lambda-\lambda_{H\beta})^{-2}$.
A saturation behaviour is seen 
near the line center of H$\beta$ at which the Raman scattering optical
depth becomes larger than unity. The strength of the Raman scattered 
He~II~$\lambda$~4850 also increases as $N_{HI}$. 

In Fig.~7, in order to illustrate the centre shift phenomenon clearly,
we overplot profiles of the Raman scattered He~II~$\lambda$~4850
obtained for various $N_{HI}$ as we increase $N_{HI}$ starting from
$N_{HI}=2\times 10^{20}{\rm\ cm^{-2}}$ by a step of $\Delta N_{HI}
=4\times 10^{20}{\rm\ cm^{-2}}$. Here, we use the same incident
radiation as in Fig.~6. Because the strength of the 4850 feature
increases as $N_{HI}$, stronger profiles correspond to higher $N_{HI}$.
It is immediately seen in this figure that the amount of centre shift
decreases as $N_{HI}$.

\begin{figure}
 \vspace{10pt}
 \hspace{-10pt}
 \epsfig{file=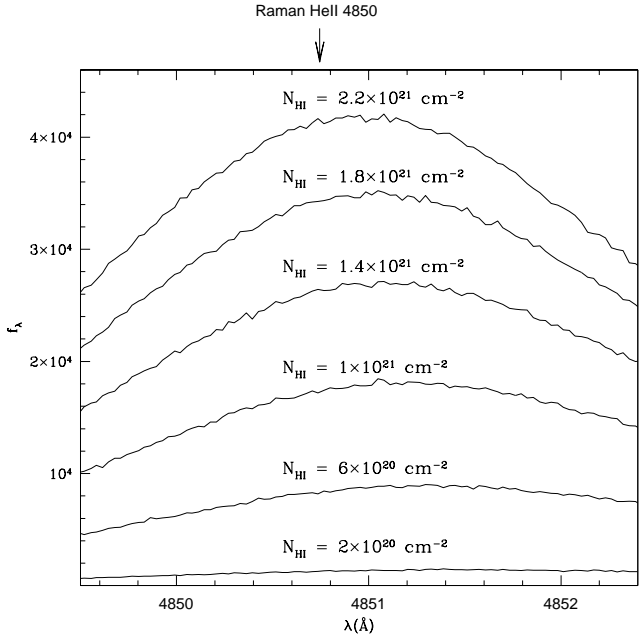, height=9cm, width=9cm}
 \caption{
The profiles of He~II~$\lambda$~4850 feature obtained for various
column densities overplotted.
The incident He~II~$\lambda$~972 is assumed
to be given by a single Gaussian with a width $\Delta\lambda = 0.076{\rm\
\AA}$ and the neutral scattering region is a plane-parallel slab
characterized by a single column density $N_{HI}$. $N_{HI}$ is
increased from $N_{HI}=2\times 10^{20}{\rm\ cm^{-2}}$ in steps of $\Delta N_{HI}
=4\times 10^{20}{\rm\ cm^{-2}}$. 
}
\end{figure}

In Fig.~8, we show the centre shift of He~II~$\lambda$~4850 feature as
a function of $N_{HI}$, treating the scattering region as a  plane-parallel slab
characterized by a single column density. As is emphasized earlier, 
we assume that
the scattering region is static with respect to the emission region
and measure the relative shift from the He~II~$\lambda$~4861.
There exist fluctuations in the result shown in Fig.~8, which is indicative
of the statistical nature of the Monte Carlo calculation. We provide
a simple fit to our data, which is given by
\begin{equation}
(\Delta\lambda/1{\rm\ \AA})
=0.73-0.015x+0.83(0.8+x)^{-0.9},
\end{equation}
where $x=N_{HI}/10^{20}{\rm\ cm^{-2}}$ is the H~I column density
in units of $10^{20}{\rm\ cm^{-2}}$.

The behaviour of the centre shift for He~II~$\lambda$~4850 
as a function of $N_{HI}$ is almost similar to that 
for He~II~$\lambda$~6545. For $N_{HI}\ge 10^{19}{\rm\ cm^{-2}}$, the
amount of centre shift decreases as $N_{HI}$ increases. We omit the
description of our results, which can be found for 
Raman scattered He~II$\lambda$~6545 in JL04.
According to our Monte Carlo calculation, the observed centre shift by
an amount $0.64{\rm\ \AA}$ is obtained when $N_{HI}=1.2\times
10^{21}{\rm\ cm^{-2}}$. 
It is particularly notable that the slope of the centre shift as
a function of $N_{HI}$ is small for high column densities, so that
the accurate determination of $N_{HI}$ may be significantly affected
by the relative motion between the He~II emission region and the
neutral scattering region.

\begin{figure}
 \vspace{12pt}
 \hspace{-10pt}
 \epsfig{file=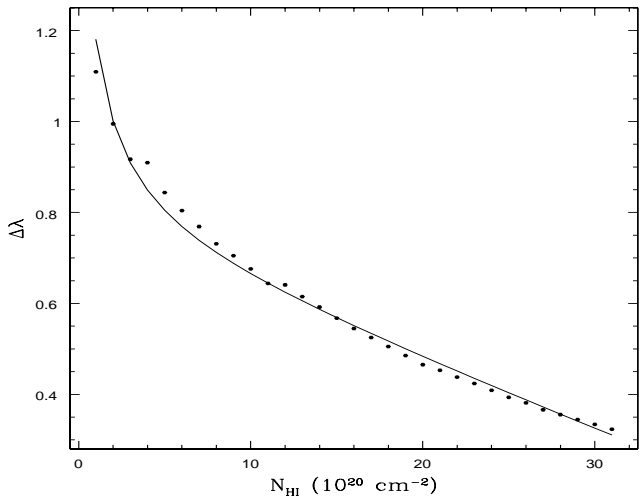, height=7cm, width=9cm}
 \caption{
The centre shift of He~II~$\lambda$~4850 feature as
a function of $N_{HI}$. The incident He~II~$\lambda$~972 is assumed
to be given by a single Gaussian with a width $\Delta\lambda = 0.076{\rm\
\AA}$ and the neutral scattering region is a plane-parallel slab
characterized by a single column density $N_{HI}$.  We give a simple fit
to our data by $(\Delta\lambda/1{\rm\ \AA})
=0.73-0.015x+0.83(0.8+x)^{-0.9}$, 
where $x=N_{HI}/10^{20}{\rm\ cm^{-2}}$ is the H~I column density
in units of $10^{20}{\rm\ cm^{-2}}$.
}
\end{figure}

\subsection{Estimate of the Mass Loss Rate}

\begin{figure}
 \vspace{12pt}
 \hspace{-4pt}
 \epsfig{file=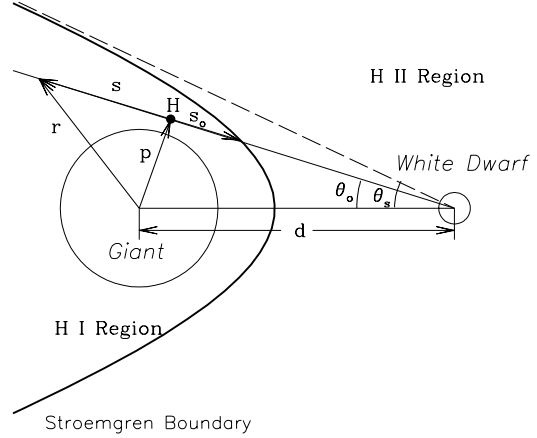, height=8cm, width=8cm}
 \caption{
A schematic diagram for the scattering geometry in
V1016~Cyg. Depending on the density structure around the giant and the
strength of the UV radiation from the white dwarf, the Str\"omgren
boundary is formed, which is defined as the loci of the ionization
front dividing the H~I region from H~II region. 
The point $H$ is the foot
of the perpendicular to a sightline from the white dwarf with
an impact parameter $p$ making an angle $\theta_o$ with the axis
connecting the giant and dwarf. 
The parameter $s$ measures the distance from $H$ to a given point along 
the sightline with a distance $r$ from the giant
so that $r^2=s^2+p^2$.  By $\theta_s$ we denote the smallest angle
of the line of sight from the white dwarf making with the axis connecting
the giant and the white dwarf, for which no H~I region can be reached.
}
\end{figure}
Assuming a spherical stellar wind from the giant component in V1016~Cyg,
we may set the hydrogen number density
\begin{equation}
n(r) = {\dot M\over 4\pi \mu m_p v_\infty r^2},
\end{equation}
where $\dot M$ is the mass loss rate of the giant, $m_p$ is the proton
mass, $\mu$ is the mean molecular weight,
and $v_\infty$ is the terminal velocity of the slow stellar wind.

Depending on the density structure and the ionizing luminosity, 
the ionization structure will be determined (e.g. Taylor \& Seaquist 1984).
A schematic diagram for this situation is shown in Fig.~9, where $d$
is the binary separation.
In this figure, we denote the ionization front by the Str\"omgren boundary 
which separates the H~I region around the giant from the H~II region
formed around the white dwarf.  By $\theta_s$ we denote the smallest angle
of the line of sight from the white dwarf making with the axis connecting
the giant and the white dwarf, for which the H~I region cannot be reached.

We consider a sightline from the white dwarf to the H~I region.
The point $H$ is the foot
of the perpendicular to this sightline from the white dwarf with
an impact parameter $p$ making an angle $\theta_o$ with the axis
connecting the giant and dwarf. 
The parameter $s$ measures the distance from $H$ to a given point along 
the sightline with a distance $r$ from the giant
so that $r^2=s^2+p^2$.  
The H~I column density for this sightline with the impact
parameter $p$ is given by
\begin{equation}
N(p) =\int_{-s_0}^\infty n(r) ds
={\dot M \over 4\pi\mu m_p v_\infty p}\left[{\pi\over2}
+\tan^{-1}{s_0\over p}\right],
\end{equation}
where $s_0$ is the distance from $H$ to the Str\"omgren boundary.

Brocksopp et al. (2002) analysed  optical and radio images of V1016 Cyg 
obtained with the Hubble Space Telescope and the Very Large Array (VLA).
Adopting a distance of 2 kpc they suggest that
the projected binary separation of V1016 Cyg is $d=84\pm 2{\rm \ AU}$.  
Previous works by other researchers provide smaller binary
separations and shorter binary periods. For example, Schmid \& Schild
(2002) performed spectropolarimetry to investigate the variation of
the polarization direction which is expected to vary due to the
orbital motion. Based on their polarimetric method, they proposed that
the binary period may exceed 100 years.
Lee et al. (2003) investigated the strength of the 6545 feature in V1016~Cyg,
and suggested the opening half angle $\theta_s\simeq 60^\circ$ adopting
the distance estimate by Brocksopp et al. (2002). 

We adopt $\mu=1.4$ as suggested by M\"urset \& Nussbaumer (1994),
$v_\infty =10{\rm\ km\ s^{-1}}$.
We choose a sightline from the white dwarf with $\theta_o$ just a little bit
smaller than $\theta_s$, for which $p\simeq d \sin\theta_s\simeq 73{\rm\ AU}$.
We may identify the H~I column density $N_{HI}^0=1.2\times
10^{21}{\rm\ cm^{-2}}$ that yields the observed
centre shift with that for this sightline. We note that the exact
choice of $s_o$ affects the final result by a factor of two. However,
in the case where the Str\"omgren boundary is well approximated by a
hyperbola as noted by Schmid (1995), we believe that $s_o$ is not much
greater than $p$. Here, we just neglect this term and set
$N_{HI} \simeq {\dot M /[ 8\mu m_p v_\infty p]}$.
From this consideration, we may obtain
\begin{equation}
{\dot M}=3.6\times 10^{-7}{\rm\ M_\odot\ yr^{-1}}.
\end{equation}

Because our analysis in the previous section is based upon a single
column density medium, and each sight line toward the H~I region
as shown in Fig.~9 will be characterized by a higher H~I column density
than $N_{HI}^0$. Therefore, the contributions to the centre shift
from these sightlines will be smaller than that from the sightline
for the Str\"omgren boundary. In order to account for the
observed redward shift, we have to lower the column density, which will
lead to a lower mass loss rate. 
However, this result should be taken with much care, noting that
the estimate of $N_{HI}$
is highly sensitive to the kinematics between the emission region
and the neutral scattering region. 

\section{Discussion and Observational Ramifications}

The inaccuracy of the estimation of $N_{HI}$
can be attributed to a number of assumptions adopted in this work.
The first assumption is that the scattering region is static with
respect to the emission region. It is very plausible that the neutral
region around the giant component takes the form of the slow stellar
wind with a typical velocity of $10{\rm\ km\ s^{-1}}$. The amount of
center shift $0.64{\rm\ \AA}$ in the Raman scattered He~II~$\lambda$~4850
may also be obtained from the receding velocity $v=7.9{\rm\ km\ s^{-1}}$
of the scattering region with respect to the emission region. This
velocity scale is smaller than the slow stellar wind velocity, and
hence our conclusion may be significantly affected. This is much more
serious in the case of RR~Tel, for which a smaller wavelength shift
is concerned.  

However, it is also expected that the neutral region 
possesses both receding and approaching motions that will compensate 
the kinematic effect to some degree. A more refined analysis will
be possible if we can secure other Raman features including
He~II~$\lambda$~4340 formed blueward of H$\gamma$, because the amounts
of the centre shifts will differ if the effect is due to atomic physics. 
Currently, it is not feasible to determine the center location 
of this feature in our spectra. Spectra with longer exposures will 
be useful in establishing a consistent model.

The assumption that the scattering region is characterized by a single
column density may not be valid. A more realistic picture would be a
system consisting of spherical shells with density decreasing 
as $r^{-2}$, where $r$
is the distance from the centre of the giant. However, the Str\"omgren
boundary due to the strong UV radiation from the white dwarf region
complicates this simple picture. Therefore, a more realistic model
should be obtained by combining photoionization calculations that
should also yield correct emission line ratios (e.g. M\"urset \&
Nussbaumer 1994). 

The measurement of the centre shift provides an upper limit of the mass
loss rate with the assumption that the scattering region is static
with respect to the He~II emission region. On the other hand, 
Lee et al. (2003) provided a rough estimate of the lower bound
$\dot M =3\times 10^{-7}{\rm\ M_\odot\ yr^{-1}}$
for the mass loss rate by equating $N_{HI}$ of the sightline for
the Str\"omgren boundary with the column density corresponding to the
unit Raman optical depth. 
We tentatively conclude that the true mass loss rate may be found
between the lower and upper bounds found by JL04 and the current work.

When van Groningen (1993) reported the discovery of the Raman
scattered He~II~$\lambda$~4850 feature in his spectrum of the symbiotic nova
RR~Tel, he also noted that it exhibits a similar redward centre shift by
an amount $\Delta v =12 {\rm\ km\ s^{-1}}$ in the parent velocity scale. 
He attributed this centre shift to a relatively receding motion 
between the emission region and the scattering region. 
He adopted the atomic data different from ours, where the line center
of He~II~$\lambda$~972 appears at 972.09 \AA\ in his analysis. 

If we interpret his result using the atomic data adopted in this paper, 
the redward centre shift of the Raman scattered He~II~$\lambda$~4850
in RR~Tel amounts to $\Delta\lambda = 0.33{\rm\ \AA}$, a significantly
smaller than the value $\Delta\lambda \simeq 1{\rm\ \AA}$ he quoted.
If the scattering region in RR~Tel is static with respect to the He~II
emission region, this redward shift indicates the H~I column density
$N_{HI}\simeq 4\times 10^{21}{\rm\ \AA}$, which is considerably higher
than that in V1016~Cyg.  This example shows that the
analysis of Raman scattered features requires extremely accurate
atomic data and that Raman features can be a very sensitive probe of
the density structure around the giant component in symbiotic stars.

\section*{Acknowledgments}
We are very grateful to Kang Min Kim and the staffs responsible for BOES of BOAO.
We also thank the referee, Hans Martin Schmid, who provided helpful comments
that improved the presentation of this paper.
This work is a result of research activities of the Astrophysical 
Research Center for the Structure and Evolution of the Cosmos (ARCSEC) 
funded by the Korea Science and Engineering Foundation.

\label{lastpage}

\end{document}